\documentclass[a4paper,11pt]{article}
\usepackage{pos, physics, mathcomp, wrapfig, booktabs}
\newcommand{\emt}{$T_{\mu\nu}$ }

\title{Computation of the latent heat of the deconfinement phase transition of SU(3) Yang-Mills theory}
\ShortTitle{Latent heat of the deconfinement phase transition of SU(3) Yang-Mills}

\author*[a,b]{Luca Virzì}
\author[a, b]{Leonardo Giusti}
\author[a,b]{Mitsuaki Hirasawa}
\author[b]{Michele Pepe}

\affiliation[a]{Department of Physics "G.Occhialini", University of Milan-Bicocca,\\
  Piazza della Scienza 3, I-20126 Milano, Italy}

\affiliation[b]{INFN, Sezione di Milano-Bicocca,\\
Piazza della Scienza 3, I-20126 Milano, Italy}

\emailAdd{l.virzi1@campus.unimib.it}
\emailAdd{leonardo.giusti@mib.infn.it}
\emailAdd{mitsuaki.hirasawa@mib.infn.it}
\emailAdd{michele.pepe@mib.infn.it}

\abstract{
We investigate the thermal properties of $\mathrm{SU}(3)$ Yang-Mills theory across the deconfinement phase transition 
considering the framework of shifted boundary conditions in the temporal direction. 
By measuring the entropy density $s(T_c)/T_c^3$ on both sides of the phase transition at the critical temperature $T_c$, we can retrieve the latent heat $h$.
Additionally, we compute $h$ from the discontinuity in the trace anomaly of the energy-momentum tensor.
Simulations are performed at five different values of the lattice spacing, allowing us to extrapolate the results to the continuum limit.
The two observables produce compatible results, giving the combined estimate $h = 1.175(10)$ in the continuum limit, achieving a precision of about 1\%. 
Moreover, we determine the critical temperature in physical units with permille accuracy, yielding $T_c \sqrt{t_0} = 0.24915(29)$.
These results allow us to connect the confined and the deconfined phases with precision, and we present an improved computation of 
the Equation of State across the phase transition for temperatures between $0$ and $3.4 T_c$.
}

\FullConference{The 41st International Symposium on Lattice Field Theory (LATTICE2024)\\
 28 July - 3 August 2024\\
Liverpool, UK\\}

\begin{document}
\maketitle

\section{Introduction}

Insights about the thermal properties of Quantum Chromodynamics (QCD) can be derived from a detailed 
investigation of its gauge sector, described by the $\mathrm{SU}(3)$ Yang-Mills theory.
 At variance with QCD that smoothly connects low and
 high temperatures, $\mathrm{SU}(3)$ Yang-Mills theory is characterized
 by a deconfinement phase transition. While the
 first order nature of the transition is well established, the
 latent heat $h$ released during the transition is currently known with a precision
 of a few percent \cite{Bors_nyi_2014,Shirogane_2016,Shirogane:2020muc,Borsanyi:2022xml}. 
 The latent heat is defined as the discontinuity in the energy density or, equivalently, in the entropy density, 
at the transition point between the confined and the deconfined phases. 
 Within the conventional approach, $h$ is
 retrieved from
 the discontinuity in the trace anomaly of the energy-momentum tensor $T_{\mu\nu}$ (EMT) 
 at criticality. Notice that the trace anomaly is affected by an ultraviolet divergence 
 related to the mixing with the identity operator, which anyway cancels out 
 when computing its gap at the critical point. 
 In this study we consider an alternative approach, where the latent heat is obtained from the discontinuity in the entropy 
 density at the critical temperature. The implementation of shifted boundary 
 conditions \cite{Giusti_2011,Giusti:2010bb,Implications} represents a convenient framework where the thermal
 features can be studied very efficiently
 by Monte Carlo simulations \cite{Giusti_2014,Giusti_2017}. 
 Finally, we present a computation of the Equation of State of the $\rm{SU}(3)$ Yang-Mills theory
 across the deconfinement phase transition \cite{Giusti:2025fxu}, updating and complementing the results obtained in Ref. \cite{Giusti_2017}.
 The determination of the Equation of State across the critical point provides a deeper
 understanding of the thermal features and of the phase transitions of a strongly interacting gauge theory, which may 
 be relevant in the context of dark matter formation in the early Universe \cite{Boddy_2014, Soni_2016, Laine_2024}.

\section{General setup and strategy}
We formulate the $\mathrm{SU}(3)$ Yang-Mills theory on a (3+1)-dimensional lattice with size $L^3 \cross L_0$ and lattice spacing $a$. Gauge fields are represented by link variables $U_\mu(x) \in \mathrm{SU}(3)$, 
 which satisfy periodic boundary conditions in the spatial direction and shifted boundary conditions along the temporal one:
\begin{equation}
    U_\mu(L_0, \vb*{x}) = U_\mu(0, \vb*{x} - L_0 \vb*{\xi}),
\end{equation}
where $L_0 \vb*{\xi}$ is a vector with integer components in lattice units. The Wilson gauge action is given by:
\begin{equation}
S[U] = \beta \sum_{x} \sum_{\mu < \nu} \left[ 1 - \frac{1}{3} \Re \Tr[U_{\mu\nu}(x)] \right],
\end{equation}
where the trace is taken over the color index, $\beta = 6/g_0^2$ is the inverse of the bare coupling $g_0^2$ and $U_{\mu\nu}(x)$ is the plaquette field
\begin{equation}
    U_{\mu\nu}(x) = U_\mu(x)U_\nu(x + a\hat{\mu})U_\mu^\dagger(x + a\hat{\nu})U_\nu^\dagger(x),
\end{equation}
with $x$ being the space-time coordinate, $\mu, \nu = 0, \dots, 3$ and $\hat{\mu}$ is the unit vector along the corresponding direction.
In the thermodynamic limit, due to the invariance under Poincaré transformations \cite{Implications}, the system with shifted boundary conditions
is equivalent to a system with periodic boundary conditions but temporal extent given by $L_0\sqrt{1 + \vb*{\xi^2}}$, which specifies the inverse
temperature $T^{-1}$. 
In the framework of shifted boundary conditions, the
off-diagonal elements of the energy-momentum tensor are
no longer all vanishing and we can define the entropy
density $s(T)$ at the temperature $T$ through the 
expectation value of the space-time components of the EMT $\langle T_{0k} \rangle_{\vb*{\xi}}$ \cite{Implications}
\begin{equation}\label{eq:entropy}
    \frac{s(T)}{T^3} = - \frac{(1 + \vb*{\xi^2})}{\xi_k} \frac{Z_T \langle T_{0k} \rangle_{\vb*{\xi}}}{T^4},
\end{equation}
where the energy-momentum tensor on the lattice is defined as follows \cite{Caracciolo:1989pt}
\begin{equation}
    T_{\mu\nu} = \frac{\beta}{6}\left\{F_{\mu\alpha}^a F_{\nu\alpha}^a - \frac{1}{4}F_{\alpha\beta}^a F_{\alpha\beta}^a\right\}.
\end{equation}
The field strength tensor $F_{\mu\nu}(x) = F_{\mu\nu}^a(x)T^a$ on the lattice is given by
\begin{equation}
    F_{\mu\nu}^a(x) = - \frac{i}{4a^2} \Tr\left\{ \left[ Q_{\mu\nu}(x) - Q_{\nu\mu}(x) \right] T^a \right\},
\end{equation}
with $T^a \in \mathfrak{su}(3)$ being the generators of the group $\mathrm{SU}(3)$, normalized as $2 \Tr \left[ T^a T^b \right] = \delta^{ab}$.
The clover field $Q_{\mu\nu}(x)$ is defined as the sum of the four coplanar plaquettes resting on the lattice site $x$:
\begin{equation}
    Q_{\mu\nu}(x) = U_{\mu\nu}(x) + U_{\nu - \mu}(x) + U_{-\mu-\nu}(x) + U_{-\nu\mu}(x),
\end{equation}
with the minus sign standing for the negative direction.

The lattice regularization breaks explicitly 
the invariance of the theory under translations 
and rotations, hence the energy-momentum tensor on 
the lattice is not a conserved quantity. 
In order to have a lattice definition of the EMT that approaches the continuum one when 
$a/L \to 0$, \emt must be multiplicatively renormalized
\cite{Caracciolo:1989pt} and $Z_T(g_0^2)$ is the 
renormalization constant of the sextet component of the tensor.
The renormalization constant in the pure $\rm{SU}(3)$ gauge theory has been computed non-perturbatively in 
\cite{Giusti_2017,Giusti_2015}.

The latent heat is defined as the difference in the energy 
density - or, equivalently, in the entropy density - between the coexisting phases 
at criticality, hence the first step of this study consists
in the determination of the critical temperature $T_c$.
Then we perform two separate Monte Carlo simulations at $T_c$, where we compute the entropy density in the 
confined and deconfined phase, $s(T^-_c)$ and
$s(T^+_c)$ respectively, and extrapolate their value to the continuum limit.
The latent heat $h$ is then given by
\begin{equation}\label{eq:lh}
    h = \frac{\Delta s(T_c)}{T_c^3} = \frac{s(T^+_c) - s(T^-_c)}{T_c^3}.
\end{equation} 
 Alternatively, we can compute $h$ from the discontinuity in the trace anomaly $A(T)$ of the EMT at the 
 two sides of the phase transition. Using the definition related to the action density \cite{Boyd_1996}, we have
 \begin{equation}\label{eq:lh_ta}
  h = \frac{\Delta A(T_c)}{T_c^4} = \frac{d\beta}{d \log(a)} \frac{a^4}{\beta L_0 L^3} \frac{\langle S(T_c^+) \rangle - \langle S(T_c^-) \rangle}{T_c^4},
 \end{equation}
 where the dependence of the lattice spacing on the bare gauge coupling can be found in \cite{Giusti:2018cmp}.

\section{Determination of the critical coupling}
We have performed Monte Carlo simulations at five different values
of the lattice spacing, corresponding to systems with temporal extension
$L_0/a = 5, 6, 7, 8$ and 9. In all cases we adopted
shifted boundary conditions with shift vector $\vb*{\xi} = (1, 0, 0)$, 
since for this value small lattice artifacts have been previously observed for 
$\langle T_{0k} \rangle_{\vb*{\xi}}$ \cite{Implications,Giusti_2014,Giusti_2015}.
Gauge configurations have been generated with 
the standard overrelaxed Cabibbo-Marinari algorithm \cite{Cabibbo:1982zn,Creutz:1980zw}. 
Close to criticality, Monte Carlo simulations show long autocorrelation times, thus we need large
statistics - $\mathcal{O}(10^6)$ configurations - to obtain accurate numerical estimates.
Various observables can be considered to define the critical coupling $\beta_c$,
but they are all equivalent in the thermodynamic limit: in this study 
we use the quantity proposed in \cite{Francis_2015,Potts}, which shows
a rapid convergence to the infinite volume value of $\beta_c$.
In correspondence of first
order phase transitions, an accurate determination of the critical coupling 
by numerical simulations at finite volume relies on a proper sampling of the coexisting phases, which can be 
obtained when many tunneling events occur in the Monte Carlo history. The probability of a tunneling event decays exponentially as the spatial size of
the system increases, thus a fast convergence to the thermodynamic limit enables us to work with lattices with moderate spatial volumes. 

At the deconfinement phase transition the $\mathbb{Z}_3$ center symmetry of the $\mathrm{SU}(3)$ Yang-Mills theory 
gets spontaneously broken: in the confined (cold) phase we have a single 
vacuum, while in the deconfined (hot) phase there are 3 degenerate vacua.
The expectation value of the Polyakov loop $\langle \Phi \rangle$ is an order parameter of this transition, and it
can be used to characterize the different phases.
When shifted boundary conditions are taken into account, the $\mathbb{Z}_3$ center symmetry of the theory is unaffected. 
However, the usual definition of the Polyakov loop needs to be modified 
in order to guarantee gauge invariance, but it remains charged under that symmetry.
We define the modified Polyakov loop for $\xi = (1, 0, 0)$ as follows
\begin{equation}\label{eq:loop}
    \Phi(\vb*{x}) = \prod_{n = 0}^{L_0/a - 1} U_0(na, \vb*{x}) \prod_{n = 0}^{L_0/a - 1} U_1(0, \vb*{x_n}),
\end{equation}
where $\vb*{x_n} = \vb*{x} - (L_0 - na)\vb*{\xi}$. 
It is useful to consider also the $\mathbb{Z}_3$ projection
\begin{equation}
    \phi = \Re{\frac{1}{(L/a)^3} \left( \sum_x \frac{1}{3} \Tr[\Phi(\vb*{x})] \right) \bar{z}},
\end{equation}
with $z$ being the $\mathrm{SU}(3)$ center element closest to the spatial average of $\Tr[\Phi(\vb*{x})]/3$. 
Close to the critical point, the probability distribution of $\phi$ exhibits two peaks corresponding to the confined and deconfined phases, 
separated by a minimum located at $\phi_0$. In a finite volume, however, the phase categorization of a field configuration
is somehow conventional: if we label with $\omega_{\rm c}$ and $\omega_{\rm d}$, respectively, the probability of finding the system 
in the confined or in the deconfined phase, the phase transition takes place when \cite{Potts,Francis_2015}
\begin{equation}\label{eq:criticality}
    \omega_{\rm d} = 3\omega_{\rm c},
\end{equation}
where the 3 factor on the r.h.s. is due to the 3-fold degeneracy of the vacuum in the deconfined phase.
The numerical estimation of these statistical weights provides a reliable result only if both phases have 
been correctly sampled, namely if many tunneling events occurred during the simulation.
\begin{figure}[t]
  \centering
  \begin{minipage}{0.485\hsize}
    \centering
    \includegraphics[width=\hsize]{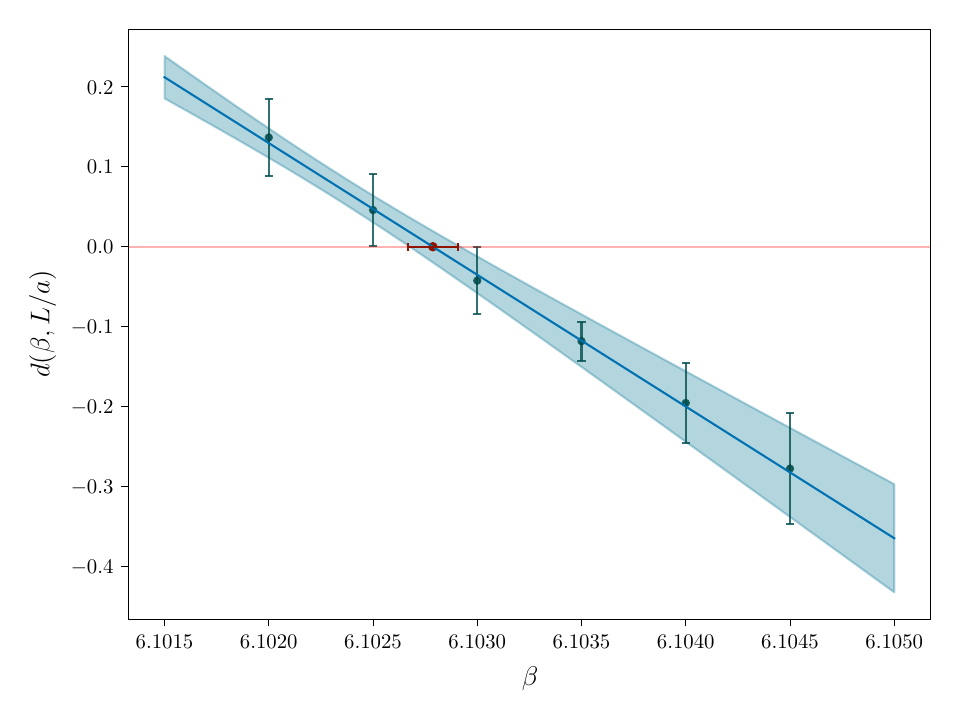}
    \caption{Dependence of $d(\beta, L/a)$ on $\beta$ for $L_0/a = 6$ and $L/a = 48.$ The critical coupling $\beta_c(L_0/a, L/a)$ - red horizontal point - is given by the zero-crossing of the linear fit function.
    The blue band represents a fit of the data.}
    \label{fig:linear}
  \end{minipage}
  \hfill
  \begin{minipage}{0.485\hsize}
    \centering
    \includegraphics[width=\hsize]{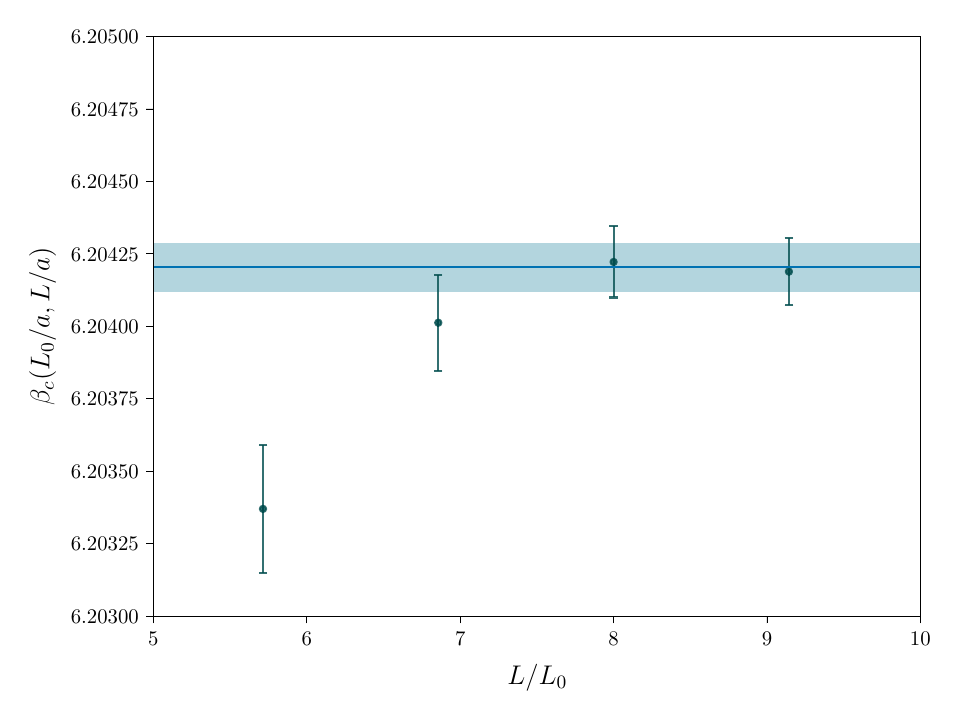}
    \caption{Dependence of $\beta_c(L_0/a, L/a)$ on $L/L_0$ for $L_0/a = 7$. The band represents the estimate of $\beta_c(L_0/a)$ coming from the weighted average of the data in the thermodynamic limit plateau, i.e. obtained at the two largest values of $L/L_0.$}
    \label{fig:inf_vol} 
  \end{minipage}
\end{figure}
Following \cite{Francis_2015}, we can then define the two probabilities as:
\begin{equation}\label{eq:weights2}
  \omega_c (\beta, L/a) = \langle \theta(\phi_0 - \phi) \rangle, \hspace{5mm} \omega_d (\beta, L/a) = \langle \theta(\phi - \phi_0)  \rangle,
\end{equation}
where $\theta$ is the Heaviside step function, and the local minimum $\phi_0$ is found by means of a quartic fit in the region 
between the two peaks. Notice that with this approach the estimation of the statistical weights should be 
affected by finite-size effects which decay exponentially with the spatial extent of the lattice \cite{Francis_2015}.
Hence on a finite lattice one can consider the quantity
\begin{equation}\label{eq:weights1}
    d(\beta, L/a) = \frac{3\omega_c(\beta, L/a) - \omega_d(\beta, L/a)}{3\omega_c(\beta, L/a) + \omega_d(\beta, L/a)},
\end{equation}
which, by definition, vanishes when the criticality condition (\ref{eq:criticality}) is met.
The critical coupling $\beta_c$ is then retrieved through a finite-size scaling study: for a fixed lattice geometry, we compute the 
parameter $d(\beta, L/a)$ and  
define $\beta_c(L_0/a, L/a)$ from the zero-crossing of $d(\beta, L/a)$ through 
a linear interpolation of the collected data, see Figure \ref{fig:linear} for an example with $L_0/a = 6$ and $L/a = 48$.
Finally, the critical coupling is obtained 
by extrapolating $\beta_{\textrm{c}}(L_0/a, L/a)$ to the infinite volume limit, i.e. $\beta_c(L_0/a) = \lim_{L/a \to \infty} \beta_c(L_0/a, L/a)$.
In Figure \ref{fig:inf_vol} we report an example for $L_0/a = 7$. 
Data show a rapid convergence to the infinite spatial volume limit, and the final result is estimated through a weighted average of the data points in the thermodynamic limit plateau.
The estimated values of the critical coupling for $L_0/a = 5, 6, 7, 8$ and $9$ are reported in Table \ref{tab:inf_beta}. 
We can now express the critical temperature $aT_c = (a/L_0)/\sqrt{2}$ in physical units with high precision.
We consider the gradient-flow time $t_0$ \cite{L_scher_2010} to set the scale and we refer to the results in \cite{Giusti:2018cmp} to relate $t_0/a^2$ to the gauge coupling $\beta$. 
Lattice artifacts on $T_{\textrm{c}}\sqrt{t_0}$ are very small, and a linear extrapolation in $(a/L_0)^2$ to the continuum gives 
\begin{equation}\label{eq:crit_temp}
  T_{\textrm{c}} \sqrt{t_0} = 0.24915(29),
\end{equation}
with a $1\tcperthousand$ final precision, mainly due to the uncertainty on the relation between $t_0$ and $\beta$. This estimate is in agreement with the analogous
result in \cite{Borsanyi:2022xml} in units of the scale $w_0$, once that the relation between the two different scales is taken into account \cite{Knechtli_2017}.

\section{Latent heat}
\begin{figure}
  \centering
  \begin{minipage}[t]{0.485\hsize}
    \centering
    \includegraphics[width=\hsize]{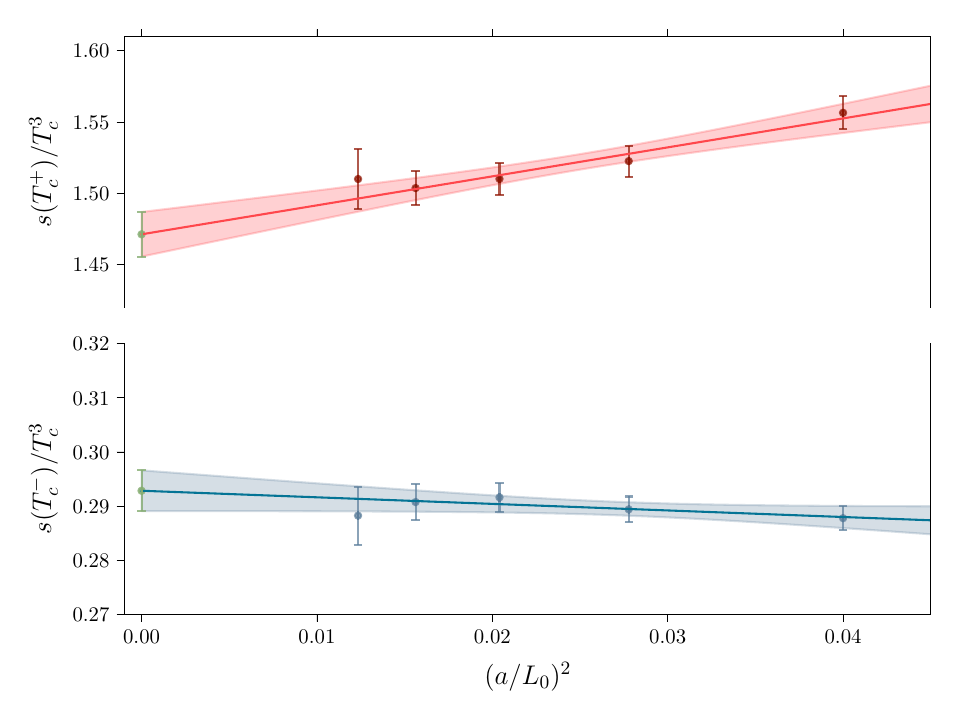}
    \caption{Extrapolation to the continuum limit of the entropy density: in the upper panel we show the deconfined phase, $s(T_c^+)/T_c^3$, while in the lower panel we show the confined phase, $s(T_c^-)/T_c^3$. The colored bands represent linear fits of the data.}
    \label{fig:entropy-fit}
  \end{minipage}
  \hfill
  \begin{minipage}[t]{0.485\hsize}
    \centering
    \includegraphics[width=\hsize]{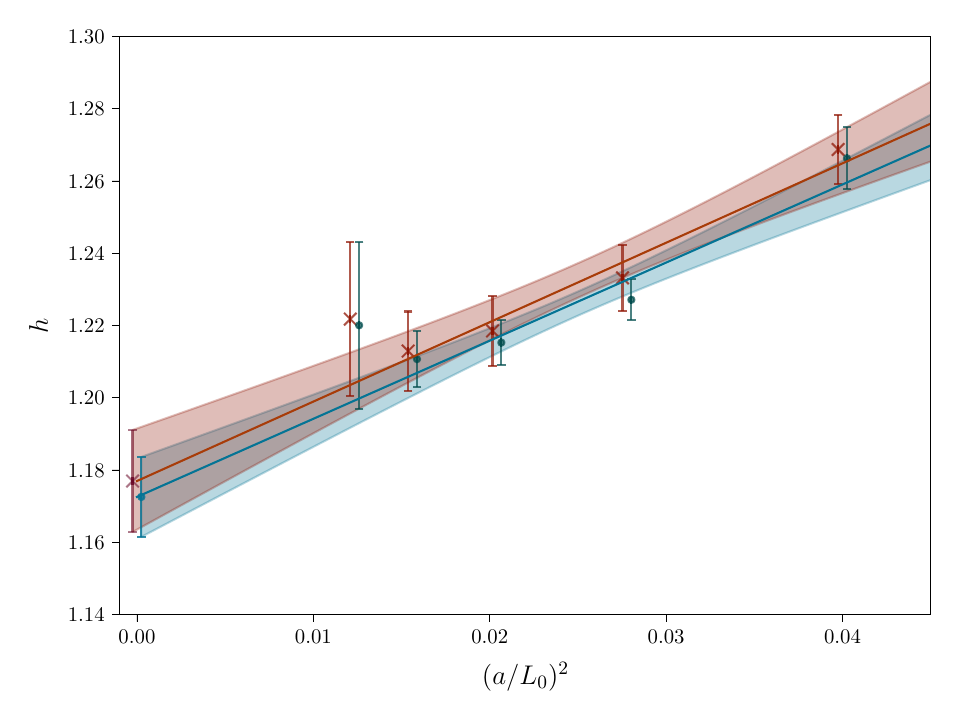}
    \caption{Extrapolation to the continuum limit of the latent heat $h$.
    Crosses (red) and dots (blue) represent, respectively, the data retrieved from $\Delta s$ and $\Delta A$.
    The bands represent linear fits of the data. Data have been slightly displaced to improve readability.}
    \label{fig:lh_cont} 
  \end{minipage}
\end{figure}

We compute the latent heat of the $\rm{SU}(3)$ Yang-Mills theory from the discontinuity between the confined and deconfined phases at $T_c$ of both
the entropy density and the trace anomaly.
We unambiguously perform Monte Carlo simulations in either phase by considering 
lattices with very large spatial extent, so that the probability of a tunneling event to the other phase is negligible.
The deconfined system is prepared by choosing an ordered field configuration as our initial 
condition, while the confined system is prepared by using a field configuration thermalized at $T < T_c$ as the starting point. 
Results are extrapolated to the continuum limit by considering values of $\Delta s(T_c)/T_c^3$ and $\Delta A(T_c)/T_c^4$ computed at 
five different lattice spacings $L_0/a = 5, 6, 7, 8$ and 9, at the critical couplings reported in Table \ref{tab:inf_beta}.
In order to avoid additional finite-size effects due to shifted boundary conditions, the aspect ratio of each lattice must be an even integer number for $\vb*{\xi} = (1, 0, 0)$ - as it was pointed out in \cite{Implications} - 
so we choose $L/a = 288$ for $L_0/a = 6, 8, 9$, while for $L_0/a = 5, 7$ we use $L/a = 280$.
Moreover, finite volume effects are negligible with respect to the statistical uncertainty, given the large spatial volumes we employed \cite{Giusti_2017}.
In Figure \ref{fig:entropy-fit} we display the dependence on $(a/L_0)^2$ of the entropy density 
in the confined phase (lower panel) and in the deconfined phase (upper
panel), alongside with their continuum limit extrapolations. Lattice artifacts are small in the cold phase
and moderate in the hot one, and data are well described by a linear fit in both cases, which give 
\begin{equation}
  \frac{s(T^-_c)}{T_c^3} = 0.2928(38), \hspace*{5mm} \frac{s(T^+_c)}{T_c^3} = 1.471(16).
\end{equation}
Consequently, as Figure \ref{fig:lh_cont} shows (red crosses), the lattice artifacts for $h$
are moderate and data follow a linear behaviour in $(a/L_0)^2$. 
In Figure \ref{fig:lh_cont} we also show the
continuum extrapolation of the latent heat measured from the discontinuity in the trace anomaly (blue dots), and overall we obtain
\begin{equation}
  h = \frac{\Delta s(T_c)}{T_c^3} = 1.177(14), \hspace*{5mm} h = \frac{\Delta A(T_c)}{T_c^4} = 1.173(11),
\end{equation}
which result in the combined estimate $h = 1.175(10)$.
\section{Equation of State}\label{sec:eqofst}
\begin{wraptable}{r}{4.5cm}
  \vspace*{-1.0cm} 
  \centering
  \begin{tabular}{cc}\\\toprule  
      $L_0/a$ & $\beta_{c}(L_0/a)$ \\
      \midrule
      5 & 5.99115(4) \\
      6 & 6.10285(8) \\
      7 & 6.20420(8) \\
      8 & 6.29626(12) \\
      9 & 6.38017(16) \\ \bottomrule
  \end{tabular}
  \caption{Infinite volume limit of the critical coupling $\beta_c$ for different lattice sizes.}\label{tab:inf_beta}
  \end{wraptable} 
 Using the data for the critical couplings $\beta_c(L_0/a)$ listed in Table \ref{tab:inf_beta} and the 
 determination of $T_c$ in units of $\sqrt{t_0}$ in Eq. (\ref{eq:crit_temp}), we have evaluated the entropy density and the pressure of the system
 for temperatures between 0 and 3.433 $T_c$.
Monte Carlo simulations have been carried out on lattices 
with spatial size $L/a = 280$ for $L_0/a = 5$ and 7 and with $L/a = 288$ for $L_0/a = 6$ and 8, and all the results for the entropy density 
are reported in Table 3 of \cite{Giusti:2025fxu}.

\begin{figure}
  \centering
  \begin{minipage}[t]{0.485\hsize}
    \centering
    \includegraphics[width=\hsize]{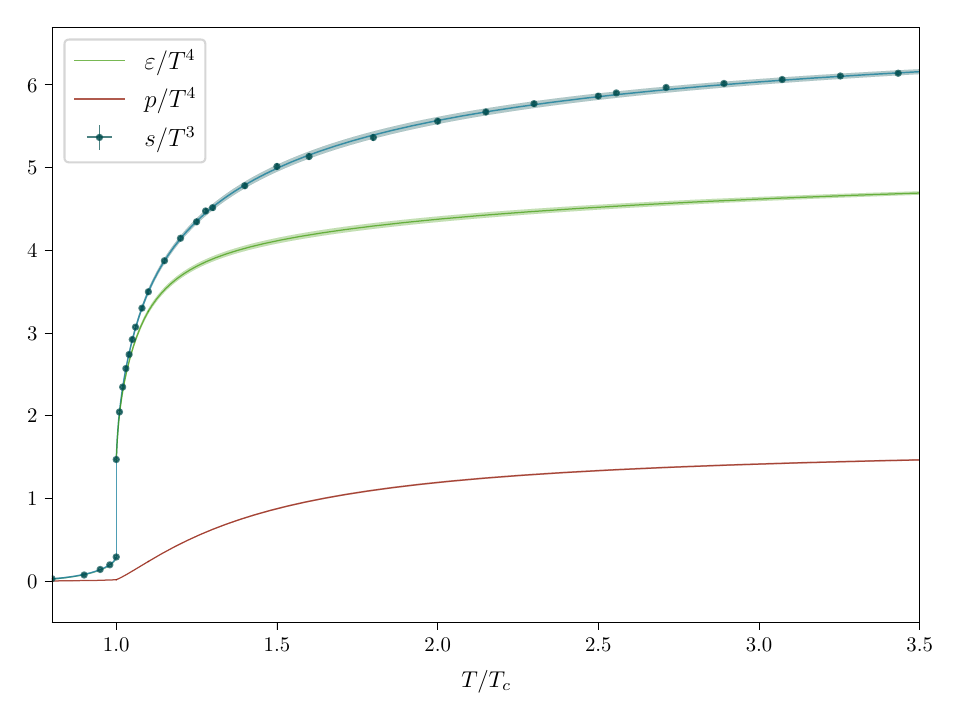}
    \caption{Dependence of the potentials $s(T)/T^3$ (blue, top), $\varepsilon(T)/T^4$ (green, middle) and $p(T)/T^4$ (red, bottom) on $T/T_c$ in the range $T/T_c \in [0.8, 3.433]$. Bands represent the interpolating functions described in Section \ref{sec:eqofst}.}
    \label{fig:eos-eps}
  \end{minipage}
  \hfill
  \begin{minipage}[t]{0.485\hsize}
    \centering
    \includegraphics[width=\hsize]{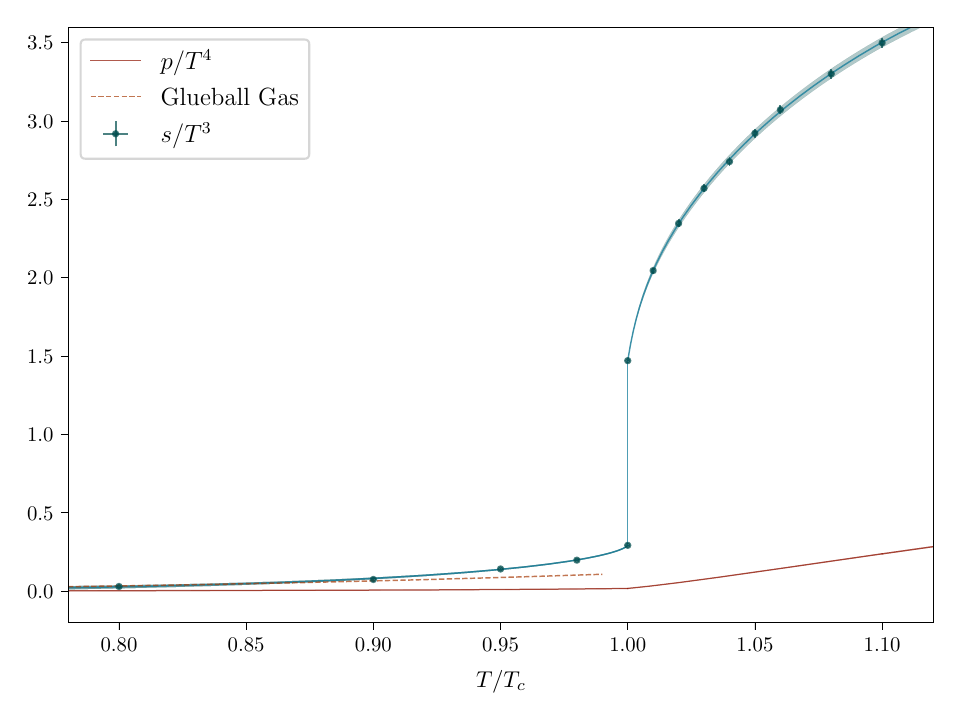}
    \caption{Detail of the dependence of $s(T)/T^3$ and $p(T)/T^4$ in the range $T/T_c \in [0.8, 1.1]$, alongside
    the non-interacting glueball gas expectation (red-dashed curve).}
    \label{fig:eos-zoom} 
  \end{minipage}
\end{figure}

For temperatures $T/T_c \in [1, 3.433]$ we consider a Padé fit 
\begin{equation}
  \frac{s(T)}{T^3} = \frac{s_1 + s_2 w + s_3 w^2}{1 + s_4 w + s_5 w^2},
\end{equation}
where $w = \log(T/T_c)$. At the lowest investigated temperature, $T/T_c = 0.80$, data agrees with the expectation coming from
a model based on a gas of non-interacting relativistic glueballs (see the red-dashed curve in Fig. \ref{fig:eos-zoom}), where the pressure given by
a particle of mass $m$ at temperature $T$ is: 
\begin{equation}
  p(m, T) = (2J + 1) \frac{(mT)^2}{2\pi^2} \sum_{n = 1}^{\infty} \frac{1}{n^2} K_2 \left(n \frac{m}{T} \right),
\end{equation}
where $K_2$ is a modified Bessel function, and $J$ is the spin of each particle. 
We have considered the four lightest glueball states $0^{++}, 0^{-+}, 2^{++}, 2^{-+}$ 
with mass lower than $2m_{0^{++}}$. Their masses, in units of $\sqrt{\sigma}$, are respectively 3.405(21), 5.276(45), 4.894(22), 6.32(9) \cite{Athenodorou_2021}.
These values can be expressed in units of $T_c$ through the relation $T_c/\sqrt{\sigma} = 0.6462(30)$ \cite{Lucini_2004} with the string tension $\sigma$.
At higher temperatures heavier states contribute to the entropy density and the
Hagedorn spectrum \cite{Hagedorn:1965st} describes how this can happen. Overall, a phenomenological fit 
$(a_0 + a_1 t^{1/3} + a_2 t^{2/3} + a_3 t)$, with $t = (1 - T/T_c)$, gives a good interpolation of our numerical results for $s(T)/T^3$ 
in the range $T/T_c \in [0.8, 1]$. Once that a convenient parametrization of the entropy density has been found,
we can determine the pressure $p(T)$ by integrating $s(T)$ in the temperature.
For temperatures in the range $T/T_c \in [0.8, 1]$ we can represent the pressure with a cubic fit function
\begin{equation}
  \frac{p(T)}{T^4} = p_0 + p_1 t + p_2 t^2 + p_3 t^3,
\end{equation}
while above $T_c$ we parametrize the pressure with a Padé interpolant
\begin{equation}
  \frac{p(T)}{T^4} = \frac{p_1 + p_2 w + p_3 w^2}{1 + p_4 w + p_5 w^2}.
\end{equation}
Once that the entropy density and the pressure are known, we can retrieve the energy density from the thermodynamic relation
$Ts = \varepsilon + p$. In Figure \ref{fig:eos-eps} we show the dependence of these thermodynamic potentials on the temperature in the range
$T/T_c \in [0.8, 3.433]$.
In Figure \ref{fig:eos-zoom} we show more closely the behaviour of the entropy density and of the pressure near the phase transition,
alongside the non-interacting glueball gas prediction. All the reported results have been extrapolated to the continuum, and the details of each parametrization 
can be found in \cite{Giusti:2025fxu}.
With this computation of the Equation of State of the $\textrm{SU}(3)$ Yang-Mills theory across the deconfinement
phase transition we complement the results obtained in Ref. \cite{Giusti_2017}.

\section{Conclusions}
We explored the thermal properties of the $\rm{SU}(3)$ Yang-Mills theory across the deconfinement phase transition
in the framework of shifted boundary conditions.
We determined the critical temperature $T_c$ in physical units with a permille precision,
yielding $T_c\sqrt{t_0} = 0.24915(29)$, then we computed the latent heat from the discontinuity of both the entropy density 
and the trace anomaly at criticality.
A combined estimate in the infinite spatial volume and continuum limit gives $h = 1.175(10)$, with a precision of about 1\%.
This result is in tension with the current best estimate \cite{Borsanyi:2022xml}
of $h$, while it is consistent with the result quoted in \cite{Shirogane:2020muc} for a fixed aspect ratio.
Finally, we performed a precise determination of the Equation of State of the $\rm{SU}(3)$ Yang-Mills theory across the deconfinement phase transition,
complementing the results of Ref. \cite{Giusti_2017}.

\acknowledgments{We acknowledge YITP in Kyoto University 
for granting us access to the supercomputer Yukawa-21.
We also thank CINECA for providing us with a very generous 
access to Leonardo during the early phases of operations
of the machine and for the computer time allocated thanks to
CINECA-INFN and CINECA-Bicocca agreements. The R\&D
has been carried out on the PC clusters Wilson and Knuth at
Milano-Bicocca. This work is (partially) supported by ICSC -
Centro Nazionale di Ricerca in High Performance Computing,
Big Data and Quantum Computing, funded by European Union
- NextGenerationEU. We wish to thank Leonardo Cosmai for
useful discussions and Isabella Leone Zimmel for her 
contribution during the early stages of this collaboration.}


\nocite{*}

\bibliographystyle{JHEP.bst}
\bibliography{bibliography.bib}

\end{document}